\documentstyle[12pt]{article}
\newcommand{\be}{\begin{equation}}
\newcommand{\ee}{\end{equation}}
\newcommand{\ba}{\begin{eqnarray}}
\newcommand{\ea}{\end{eqnarray}}
\newcommand{\baa}{\begin{eqnarray*}}
\newcommand{\eaa}{\end{eqnarray*}}
\newcommand{\lab}[1]{\label{#1}}

\newcommand{\bhat}{\hat{\beta}}

\begin{document}
{\pagestyle{empty}
\rightline{SC-94-}
\rightline{NWU-5/94}
\rightline{July 1994}
\vskip 2.5cm
{\renewcommand{\thefootnote}{\fnsymbol{footnote}}
\centerline{\large \bf Comparative Study of Multicanonical and
       Simulated Annealing}

\vskip 0.5cm
\centerline{\large \bf Algorithms in the Protein Folding Problem}
}
\vskip 3.0cm

\centerline{Ulrich H.E.~Hansmann$^1$ and Yuko Okamoto$^2$}
\vskip 1.5cm

\centerline{$^1${\it Konrad Zuse Zentrum f{\"u}r Informationstechnik
Berlin (ZIB) }}
\centerline{{\it  10711 Berlin, Germany}}
\vskip 0.5cm

\centerline {$^2${\it Department of Physics,
Nara Women's University, Nara 630, Japan}}

\medbreak
\vskip 3.5cm

\centerline{\bf ABSTRACT}
     We compare a few variants of the recently
proposed multicanonical method
with the well known simulated annealing for the effectiveness in search
of the energy global minimum of a biomolecular system.  For this we
study in detail
Met-enkephalin, one of the simplest peptides.
We show that the new method not only
outperforms simulated annealing in the search of the energy groundstate
but also
provides more
statistical-mechanical information
about the system.
\vskip 0.1in

\vfill
\newpage}
\baselineskip=0.8cm
\section{INTRODUCTION}
Systems with frustration are commonly present in many fields of
science and engineering.  To name a few, spinglass, neural network,
protein folding, traveling salesman problem, and optimal wiring of
electric circuits are the examples of such systems.  These systems
belong to the same class of the so-called NP complete optimization
problem where the number of computing steps required to solve the
problem increases faster than any power of the size of the system.
The problem is simply stated as follows:  Find the global minimum
out of a huge number of local minima separated by high tunneling
barriers.

In this article, we compare two effective methods,
simulated annealing \cite{SA} and multicanonical algorithms \cite{MU,MU1},
in the study of a system of frustration by taking the example of the
protein folding problem.
The prediction of the three-dimensional structure of proteins solely
from their amino
acid
sequences remains one of the long-standing unsolved problems in
biophysics
(for a recent review, see, for example, Ref. \cite{REV}).
Traditional methods such as molecular dynamics and Monte Carlo
simulations
at relevant temperatures tend to get trapped in local minima.
A now almost
classical way to alleviate this difficulty
 is simulated annealing.\cite{SA}
Simulated annealing was proposed to be used
to predict the global minimum-energy
conformations of polypeptides and proteins \cite{SA1}-\cite{SA3}
and refine protein structures from NMR and
X-ray data.\cite{Nil}-\cite{Brun2}  Since then many promising results
have been obtained.\cite{SA4}-\cite{YOS}

Recently the authors proposed another approach, the multicanonical
 method \cite{MU,MU1},  to the protein folding problem.\cite{HO} This new
ansatz
 was  successfully tested for systems with first-order phase transitions
\cite{MU1,MU2,MU2P}  and spin glasses \cite{SG1}-\cite{SG3P}
where one has to deal with
similar multiple-minima problems.
 The core of this ansatz is to perform
 Monte Carlo simulations in a {\em multicanonical}
ensemble \cite{MU1} instead of the usual (canonical) Gibbs-ensemble.
In the new  ensemble the energy is forced on a one-dimensional
random walk
 and so a simulation can overcome the energy barriers between
local minima by connecting back to the high temperature states.
 The canonical distribution for {\em any} temperature
 can  be obtained by the re-weighting
techniques.\cite{FS}
In a pilot study \cite{HO} the authors demonstrated for a small peptide
 that the new approach allows
indeed both the identification of groundstates and
 the calculation of thermodynamic quantities over a wide temperature
 range
from just one simulation.

In this paper we like to compare the performance of the multicanonical
ansatz with the older simulated annealing.
 For this test
we chose the system of Met-enkephalin, one of the simplest peptides,
since its
lowest-energy conformation for the potential energy function ECEPP/2
\cite{EC1}-\cite{EC3} is known.\cite{Enk}
As another task we
like to study the thermodynamic properties of this peptide in greater
detail.

The paper is organized as follows. In Section~2 we briefly review
simulated annealing and different variants of the multicanonical method.
In Section~3 we give the computational details. In Section~4 we present
our results for the exploration of groundstates and compare
the different optimization methods.  Later in the Section
we try to evaluate physical
quantities of the peptide over a large temperature range.
Conclusions and discussion are given in Section~5.

\section{ALGORITHMS}
\subsection{Simulated Annealing}
Today, simulated annealing \cite{SA} is one of the most widely used
methods
in solving
global optimization
problems.
 The method is based on the \lq\lq crystal
forming" process; during a simulation temperature is lowered very slowly
from a sufficiently high initial temperature
$T_0$
where the structure changes freely
with Monte Carlo updates to a \lq\lq freezing" temperature
$T_f$
where the system
undergoes no significant changes with respect to the Monte Carlo iteration.
If
the rate of temperature decrease is slow enough for the system
to stay in
thermodynamic equilibrium,
then it is ensured that the system can avoid getting trapped in
local minima and that the global minimum
will be found.
However, the simulation should be monitored closely and the annealing
should be tailored very carefully to achieve this condition.\cite{Dave2}
This causes a great increase in required computation time especially
at low temperatures, and so it is often run with a simple annealing
protocol without a careful testing of thermal equilibrium.
Hence, the relationship of the obtained conformations to the
 the calculation of thermodynamic quantities will be
hampered by uncontrolled bias.
For this reason, a certain number (which is not known {\it a priori}) of runs
are necessary to evaluate the
performance and to make sure that the obtained final conformations are close
to the global minimum.

In the present simulation
the temperature was lowered exponentially in $NSTEP$ times by setting
the inverse temperature
$\bhat = 1/k_B T$ to
\be
\bhat_n = \bhat_0 \gamma^{n-1}~,
\lab{ana}
\ee
for the $n$th temperature step.
Here,
$\bhat_0$
is the initial inverse temperature and $\gamma$
is given by
\be
\gamma =
\left( \frac{ \displaystyle \bhat_f}{\displaystyle \bhat_0}
\right) ^{\frac 1 {NSTEP-1}}~,
\lab{anb}
\ee
In the present work, for the purpose of comparison, we made a fixed
number of Monte Carlo sweeps, $N_T$, at each temperature step so that
the total number of Monte Carlo sweeps per run, $NSWEEP$, is fixed:
\be
NSWEEP = N_T \times NSTEP~.
\lab{anc}
\ee
For a fixed value of $NSWEEP$, these constants
($\bhat_0,~\bhat_f,$ and $NSTEP$) are
free parameters and have to be
tuned in such a way that the annealing process is optimized for the
specific problem.  In the present work, we set the initial temperature
$T_0$ to be 1000 K and compared the cases for a few choices of
$NSTEP$ and the final temperature $T_f$.

\subsection{Multicanonical Ensemble}
In the canonical ensemble, configurations at an inverse temperature
 $\bhat \equiv 1/k_B T$ are weighted with  the Boltzmann factor
$
w_B (E)\ =\ \exp \left( - \bhat E \right) .
$
The resulting probability
distribution is given by
\be
P_{B}(E)\ ~\propto ~n(E) w_{B}(E)~,
\lab{pb}
\ee
where $n(E)$ is the spectral density.
In the {\em multicanonical} ensemble, \cite{MU1} on the other hand,
the probability distribution is {\it defined} by the condition
\be
P_{mu} (E) ~\propto ~ n (E) w_{mu} (E) = {\rm const}.
\lab{pd}
\ee
Hence, all energies have equal
weight and
a one-dimensional random walk in energy space is realized
(when simulated with local updates), which insures
that the system can overcome any energy barrier.
Note that from Eq.~(\ref{pd}) we have
\be
w_{mu} (E) ~\propto ~n^{-1} (E)~. \lab{e3}
\ee
Unlike for the canonical ensemble the multicanonical weight factor
$w_{mu} (E)$ is
not {\it a priori} known, and one needs its estimator
for a numerical simulation. Hence, the multicanonical ansatz consists of
three steps:
 In the first step the estimator of the multicanonical weight factor is
       calculated.
 Then one performs  with this weight factor a multicanonical
simulation with high statistics.
The standard Markov process (for instance, in a Metropolis update scheme
\cite{Metro})
is well-suited for generating configurations which
are in equilibrium with respect to the multicanonical distribution.
Finally, from this simulation one can not
 only locate the energy global minimum but also obtain the
canonical distribution at any inverse temperature $\bhat$
for a wide
range of temperatures by the re-weighting technique:\cite{FS}
\be
P_B(\bhat,E) \propto w^{-1}_{mu} P_{mu} (E) e^{-\bhat E}~.\lab{erw}
\ee

The crucial point is the first step: calculating the estimator
for the multicanonical weight factor $w_{mu} (E)$. This can be
done  by the
following iterative procedure:\cite{MU,HO}
\begin{enumerate}
\item Perform a canonical Monte Carlo simulation at a sufficiently high
 temperature $T_0$. In our case we chose $T_0 = 1000$ K. The weight
factor
 for this simulation is given by
$ w(E) = e^{-\bhat_0 E} $ with $\bhat_0 = 1/k_B T_0$.
 Initialize the array $S (E)$ to zero, where $E$ is discretized
with bin width $\delta E$ ($=1$ kcal/mol in the present work).
\item Sample the energy distribution obtained in the previous simulation
as a
histogram $H(E)$ with the same bin width as in $S (E)$.
In the first iteration (step 1 above) determine
$E_{max}$ as the value near the mode where the histogram has
its maximum.  ($E_{max}$ is fixed throughout the iterations.)
Let $E_{min}$ be the lowest energy ever obtained.
For all $H(E)$ with entries greater than a minimum
value (say, 20)
and
$E_{min} \le E \le E_{max}$,
update the array $S(E)$ by
\be
S(E) = S(E) + \ln H(E)~.
\ee
\item Calculate the following multicanonical parameters $\alpha (E)$
and $\beta (E)$ from the array $S(E)$:
 \begin{equation}
 \beta (E) = \left\{ \begin{array}{ll}
		     \bhat_0 & ,~E \ge E_{max}\cr
                 \bhat_0 +    \frac{ \displaystyle S(E') -
                                          S(E)}{\displaystyle
		     E' - E} & ,~
                      E_{min} \le E < E' < E_{max}\cr
                     \beta (E_{min}) & ,~ E < E_{min}
                      \end{array} \right.
 \end{equation}
 and
 \begin{equation}
 \alpha (E) = \left\{ \begin{array}{ll}
                      0 &,~ E \ge E_{max}\cr
                      \alpha (E') + (\beta (E') - \beta (E))E' &,~ E < E_{max}
                      \end{array} \right.
 \end{equation}
 where $E$ and $E'$ are adjacent bins in the array $ S (E)$.
 \item Start a new simulation with the multicanonical weight factor
 defined by
 \begin{equation}
 w_{mu}(E) = e^{-\beta(E) E - \alpha (E)}~.\lab{ewf}
 \end{equation}
 \item Iterate the last three steps
until the obtained distribution $H(E)$ becomes reasonably
flat in the chosen energy range.
\end{enumerate}
While this method for determining the multicanonical weight factor
$w_{mu}(E)$ is quite
general, it has the disadvantage that it
requires a certain number of iterations which is not {\it a
priori} known. For the calculations in Ref.~\cite{HO} about 40 \% of the
total CPU time was spent for this part. We remark that the
above method of calculating multicanonical weights is by no means
unique. Especially it is not necessary to choose the parametrization of
Eq.~(\ref{ewf}) for the multicanonical weight factor.
However, with this parametrization and its introduction of ``effective''
temperatures $\beta(E)$
the connection to the canonical ensemble becomes very clear.

\subsection{Multicanonical Annealing}
If one is just interested in the
groundstate structure it may be worthwhile to use instead a variant of the
multicanonical
method, {\it multicanonical annealing}. \cite{LC}
 Multicanonical annealing
alleviates the above-mentioned complication of
determination of the multicanonical weight factor.  However, because
of this simplification, it
does  not allow any
calculation of thermodynamic quantities.
It was first proposed for the traveling
salesman problem, one of the classical examples of an
NP-complete optimization problem.
First applications to the protein folding problem exists, too.\cite{HO94_2}
Results  better than those obtained
by simulated annealing were reported.

In  simulated annealing,  directed moves are achieved by gradually
lowering the temperature. For multicanonical annealing, we introduce an
upper bound in energy at
the other end of the annealing direction, rejecting all
attemps beyond this bound. Annealing is achieved by moving the bound in
the annealing direction while keeping the sampling interval $\Delta E$
fixed.  Within this energy  interval the system
can move out of local minima as long as their barrier heights do not
 exceed the upper limit of the energy range.  The algorithm can be
implemented in the
 following way: \cite{LC,HO94_2}
\begin{enumerate}
\item Perform a short canonical Monte Carlo (MC) simulation at a
sufficiently high
 temperature $T_0$. Again we chose $T_0=1000$ K in the present work.
 Initialize an array $S (E)$ to zero, where $E$ is discretized
with bin width $\delta E$ ($=1$ kcal/mol in the present work).
\item Sample the energy distribution obtained in the previous
simulation as a histogram
 $H(E)$.  Let $E_{min}$ be the lowest energy ever obtained.
Calculate
\be
  S(E) = S(E) + \ln H(E) ~.
\ee
\item Update the upper bound $E_{wall}$ of the sampling interval by
  \begin{equation}
  E_{wall} = \max (E_{last},E_{min} + \Delta E)
  \end{equation}
 where $\Delta E $ is the  size of the sampling energy range and $E_{last}$
 the energy of the last configuration.
\item Calculate the following parameter $\beta_{min}$ by:
\begin{equation}
 \beta_{min} =\frac{ \displaystyle S (E_{wall}) - S (E_{min})}{E_{wall} -
E_{min}}
 \end{equation}
\item Define the new weight factor by
 \begin{equation}
 w(E) = \left \{ \begin{array}{ll}
                 0 &,~ E > E_{wall}\cr
                 e^{-S(E)} &,~ E_{min} \le E \le E_{wall}\cr
		     e^{-S(E_{min}) - \beta_{min} (E - E_{min})} &,~ E < E_{min}
                 \end{array} \right.
 \end{equation}
 and start a new simulation with this weight factor with
the last configuration of the preceding simulation as the initial
structure.
\item Iterate the last four steps till some convergence criterion is met.
\end{enumerate}

In the present work, for the purpose of comparison with other methods,
we set the number of annealing iterations to a fixed value
(instead of checking the convergence criterion).  If we call
this number $NSTEP$ and the number of MC sweeps per iteration $N_T$,
then the number of MC sweeps per run, $NSWEEP$, is given by
Eq.~(\ref{anc}).

Because of the finite interval size $\Delta E$, the
MC procedure will no longer be
ergodic and it is not possible to find the equilibrium properties of the
system. Hence, the canonical distribution cannot be reconstructed. For the
purpose of annealing this does not matter as long as one chooses the
sampling interval large enough to allow important fluctuations throughout
the annealing process. However, since ergodicity is not fulfilled, one has
to repeat the annealing process many times with different initial
configurations,
to make sure that one has found a good approximation to the
global minimum.  Conventional simulated annealing has to cope with the same
problem.  As in
 simulated annealing,
where one does not know {\it a priori} the optimal cooling schedule,
the optimal sampling interval size
$\Delta E$ is
not known {\it a priori} for multicanonical annealing and has
to be chosen
on a trial and error basis. The problem can be eased by making
the sampling
interval size itself a dynamical variable which is changed according
to the following rule in the annealing process.
We preset the sampling interval size $\Delta E$ to a given minimum value
$\Delta_m$
and double it every time when the simulation does not find a new
global-minimum candidate for three
consecutive steps. Once a better estimate
of the global minimum is found, $\Delta E$ is reset to
its initial value $\Delta_m$.

In the present work, we try to compare the performances of the above
three methods, simulated annealing, regular multicanonical algorithm,
and multicanonical annealing.

\section{COMPUTATIONAL DETAILS}
\subsection{Potential Energy Function}
Met-enkephalin has the amino-acid sequence Tyr-Gly-Gly-Phe-Met.
For our simulations the
backbone was terminated by a neutral NH$_2$-- ~group at the N-terminus
and a neutral~ --COOH group at the C-terminus as in the previous works of
Met-enkephalin.\cite{SA2,RSA4,EnkO,HO,Enk}  The potential energy function
$E_{tot}$ that we used is given by the sum of
the electrostatic term $E_{es}$, 12-6 Lennard-Jones term $E_{vdW}$, and
hydrogen-bond term $E_{hb}$ for all pairs of atoms in the peptide together with
the torsion term $E_{tors}$ for all torsion angles.
\begin{eqnarray}
E_{tot} & = & E_{es} + E_{vdW} + E_{hb} + E_{tors},\\
E_{es}  & = & \sum_{(i,j)} \frac{332q_i q_j}{\epsilon r_{ij}},\\
E_{vdW} & = & \sum_{(i,j)} \left( \frac{A_{ij}}{r^{12}_{ij}}
                                - \frac{B_{ij}}{r^6_{ij}} \right),\\
E_{hb}  & = & \sum_{(i,j)} \left( \frac{C_{ij}}{r^{12}_{ij}}
                                - \frac{D_{ij}}{r^{10}_{ij}} \right),\\
E_{tors}& = & \sum_l U_l \left( 1 \pm \cos (n_l \alpha_l ) \right),
\end{eqnarray}
where $r_{ij}$ is the distance between the atoms $i$ and $j$, and $\alpha_l$ is
the torsion angle for the chemical bond $l$.
The parameters ($q_i,A_{ij},B_{ij},C_{ij},
D_{ij},U_l$ and $n_l$) for the energy function were adopted
from ECEPP/2.\cite{EC1}-\cite{EC3} The effect of surrounding atoms of
water was
neglected and the dielectric constant $\epsilon$ was set equal to 2.
The computer code
KONF90 \cite{KONF,KONF2} was used. The peptide-bond
dihedral angles $\omega$ were fixed at the value 180$^\circ$
for simplicity,
which leaves 19 angles $\phi_i,~\psi_i$, and $\chi_i$ as
independent variables.
Because of this choice  and a different convention for the
implementation of the
ECEPP parameters (for example, $\phi_1$ of ECEPP/2 is equal to
$\phi_1 - 180^{\circ}$ of KONF90, and energies are also different by
small irrelevant constant terms),
our results slightly differ from the one in Ref.~\cite{Enk}
which were obtained by another method but can be directly compared with
Refs.~\cite{EnkO} and \cite{HO}.

\subsection{Implementation of the Algorithms}
Preliminary runs showed that all  methods  need roughly the same amount
of CPU time for a fixed
number of MC sweeps (about 15 minutes for 10,000 sweeps
on an IBM RS6000 320H). Hence, we compared the different
methods by performing
simulations with the same number of total MC sweeps.
By setting this number to 1,000,000 sweeps we tried to ensure high
statistics.
Note that the statistics for each method are therefore
2.5 times more than those
of Ref.~\cite{EnkO}.
One MC sweep updates every
dihedral angle of the peptide once.

For simulated annealing and multicanonical annealing, one problem
is to find the
optimal annealing schedules. For each choice of the annealing
parameters,
(the final temperature and number of temperature steps
for simulated annealing and the sampling
interval size and number of annealing iterations
for multicanonical annealing),
we performed
10 simulations with $NSWEEP=100,000$
(so that the total number of MC sweeps is equal to
1,000,000). Each simulation started with a different random initial
conformation.   For certain cases, we also performed 20 runs
with $NSWEEP=50,000$ in order to see the dependence of the methods
on $NSWEEP$.

For simulated annealing the
initial temperature was $T_0 =1000$ K and the temperature was
exponentially decreased
according to Eq.~(\ref{ana}).
We compared the cases for three values of
$NSTEP$ ($NSTEP=NSWEEP,$ 50, and 20) and two final temperatures
($T_f = 50$ and 1 K).  The case for $NSTEP=NSWEEP$ is the protocol of
Ref.~\cite{KONF2}, and it represents the slowest and smoothest annealing
schedule (the factor $\gamma$ in Eq.~(\ref{ana}) is equal to
$1.000029 \cdots$ for $T_f=50$ K and $NSWEEP=100,000$).
The cases for $NSTEP=50$ ($\gamma = 1.0630 \cdots$ for $T_f=50$ K)
and for $NSTEP=20$ ($\gamma = 1.1707 \cdots$ for $T_f=50$ K) are
essentially the protocol of Ref.~\cite{RSA1}.

For multicanonical annealing we divided the
100,000 MC sweeps of each run
in 10 annealing iterations of 10,000
sweeps for the case of fixed sampling interval
and in 20 steps of 5,000 sweeps for the case of dynamically changed
sampling size.  For the former case, we compared
three values of the sampling interval size $\Delta E$
($\Delta E=5,$ 10, and 15 kcal/mol), and for the
latter case, we compared
two values of the minimal
sampling size $\Delta_m$ ($\Delta_m = 5$ and 10 kcal/mol).

For regular multicanonical simulation we used the same
weight factor as that already determined by
 four iterative steps of 10,000 sweeps each in Ref.~\cite{HO}.
 All thermodynamic quantities were then  calculated from
 a production run of 9,960,000 sweeps which includes 10,000 sweeps
for thermalization. At the end of every
second sweep we stored the actual configuration and the
energy for future analysis of thermodynamic quantities.
 Note that
 the fraction of CPU time needed for calculating the
 multicanonical weight factor can now be restated to be only 4 \%
 (instead of 40 \% in Ref.~\cite{HO}).

\section{RESULTS}
\subsection{Groundstate Investigations}
First, we have to define the groundstate of enkephalin.  In
Ref.~\cite{EnkO} it was shown that with KONF90, conformations with
energies less than $-11.0$ kcal/mol have essentially the same
structure (Type A of Ref.~\cite{EnkO}).
Hence, we consider any conformation with $E<-11.0$ kcal/mol
as the groundstate configuration.

In our production run for regular multicanonical simulations we
observed 18 tunneling events where a tunneling event means that the system went
from the groundstate region (energies less than $-11.0$ kcal/mol)
to
$E_{max}$ (above which the multicanonical parameters are set to
$\alpha(E)=0$ and $\beta(E) = \hat{\beta}_0 = 1/k_BT_0$ with
$T_0 = 1000$ K)
and came back to the groundstate region.  Here, we set
$E_{max}=20$ kcal/mol.
When the system reaches $E_{max}$ region, it has to stay a certain
amount of time with which the simulation
is updated according to the Boltzmann weight for a very
high temperature
($T_0=1000$ K).  Hence, the system
will encounter enough randomness to ensure that groundstate
configurations separated by such a tunneling event are  statistically
independent.
For this reason, the number of tunneling events gives a
lower bound for the number of independent groundstate configurations found
by this method.
Table~1  summarizes our results.
For each tunneling event the estimated groundstate energy $E_{GS}$
was chosen to be the lowest energy obtained
in the corresponding cycle.  The 18 estimated groundstate structures
were all essentially
the same, by which we mean their dihedral angles ($\phi_i$, $\psi_i$,
and $\chi_i$) differ less than $\approx 10$ degrees.
{}From the table we found as a tunneling time $\tau_{tu} =
54136 \pm 8187$, measured in MC sweeps.
 This value is high for such a small
 peptide as Met-enkephalin.  This is  caused by the
 high rejection rate and small energy
 changes of
  the accepted Metropolis steps at low energy regions.  But one has
 to remember that
 the use of the multicanonical ensemble is not restricted to local
 Metropolis updates. Work is under progress
  to accelerate the method by using collective updates.

In Table~2 we list the results obtained by simulated
annealing for
the lowest energies,
the number of times a groundstate
configuration (with energy $E \le -11.0$ kcal/mol) was found, and
the average of their lowest
energy estimates.  The results are from 10 runs with $NSWEEP=100,000$
per run.
The annealing schedules were tested for three different choices of
$NSTEP$ ($=NSWEEP$, 50, and 20) and two different
choices of the final temperature $T_f$ ($=50$ K and 1 K), where
$NSTEP$ is the number
of different temperatures considered in the annealing process
(see Eqns.~(\ref{ana})--(\ref{anc})).
The first thing one can read off from the table is that the results
are similar from one annealing schedule to another (no
order-of-magnitude differences).
However, the results imply the following qualitative tendencies.
First of all, when the annealing is fast ($NSTEP=20$), the
probability of finding the global minimum decreases compared to
the case for slower annealing ($NSTEP=NSWEEP$ and 50).
Secondly, if the final temperature is too low, then
the cooling will also be fast and with a certain probability
simulated annealing will get trapped in some metastable states.
We observed this for the final
temperature $T_f=1$ K where we found either no improvement or even
worse results than those for
$T_f=50$ K.
On the other hand, if the final temperature
is too high, then the probability of finding the
global minimum is also low.
For the final temperature $T_f = 300$ K simulated annealing failed in
finding a groundstate (data not shown).  Hence, the optimal final
temperature for the present case is around 50 K.

In Table~3 we list the same quantities obtained by simulated
annealing for less number of sweeps per run ($NSWEEP=50,000$) with the
continuous annealing ($NSTEP=NSWEEP$).
The results are worse for $T_f=50$ K compared to those in
Table~2 (the probability of finding the groundstate decreasing
from 50 \% to 30 \%
by the reduction of
the number of sweeps per run).
The situation is clearer if we compare these results with
those from
100 runs of 10,000 sweeps each (data not shown).
In the latter case, we found
groundstates only
 13 times (13 \%), and the average of the lowest energy
was $<E> = -8.4 (1.7)$ kcal/mol.
Hence, the optimal choice for $NSWEEP$ is definitely more than
10,000 and probably more than 100,000.

We now discuss the results for multicanonical annealing.
In Table~4 we list
the lowest energies,
the number of times a groundstate
configuration was found, and the average of their lowest
energy estimates.  The results are from 10 runs with $NSWEEP=100,000$
per run.
They strongly depend on the choice of the parameters that rule the
annealing procedure.
In particular, one has to
choose the sampling energy interval $\Delta E$
large enough. The probability to find the
groundstate increases with $\Delta E$ from 30 \% for $\Delta E = 5$
kcal/mol to
100 \% for $\Delta E= 15$ kcal/mol.  The probability also
increases from 80 \% for minimal sampling energy interval
$\Delta_m =5$ kcal/mol to
100 \% for $\Delta_m =10$ kcal/mol for the case of dynamically
changed sampling size.

In Table~5 we list the same quantities obtained by
multicanonical
annealing with less number of sweeps per run ($NSWEEP=50,000$)
for the optimal choices of sampling sizes from Table~4.
We performed 10 annealing steps of 5,000 sweeps each for the case of
fixed sampling interval size and 20 annealing steps of 2,500 sweeps
for the case of dynamically changed sampling size.
As noted for the case of simulated annealing, the results are
worse by reducing the number of sweeps per run; the probability
of finding the groundstate decreases
from 100 \% to 90 \% and from 100 \% to 75 \% for fixed sampling
size and dynamically changed sampling size, respectively.
Hence, the optimal choice for $NSWEEP$ is
probably between 50,000 and 100,000.

We now compare the results for simulated annealing and multicanonical
annealing (Tables~2--5).
First, we observe that the
probability of finding the groundstate was 100 \% for the optimal
cases of
multicanonical annealing, while that for simulated annealing
was 50 \%.
Secondly,
the average $<E>$ is generally smaller and fluctuates less
for multicanonical annealing than
for simulated annealing.  This is true even in the cases
where both methods
found the same number
of groundstate configurations (compare the results for
$\Delta E = 10$ kcal/mol in Table~4 and those for
$T_f=50$ K in Table~2).
Hence, we conclude that
the multicanonical annealing is superior
to simulated annealing.
However, the improvement
is not as impressive as it was quoted for
the traveling salesman problem in Ref.~\cite{LC}.
The superiority of multicanonical annealing to simulated annealing
can be understood by the fact that the
sampling interval for multicanonical annealing is essentially constant,
while it shrinks with decreasing temperature for simulated annealing.
Hence,
the chance of escaping from a local minimum
is much larger for multicanonical annealing.

On the other hand, there seem to be no
significant advantage of multicanonical annealing over
a regular multicanonical algorithm.
The maximum numbers of
independent groundstate configurations found by a regular
multicanonical
simulation ($n_{GS} = 19$) and multicanonical annealing
($n_{GS} = 18$) are almost the same (see Tables~1 and 5).
Since a regular multicanonical simulation provides additional
information about the thermodynamics of the system, it seems
that this
method is the best choice.
The situation may be different for spin glasses
and the
traveling salesman problem where one always has to study different
realizations of the systems.  Namely, one has to recalculate the
multicanonical weight factor for each realization anew and
therefore it can be
computationally easier to use multicanonical annealing.

To conclude this section, we discuss another application of
multicanonical annealing: its use for the determination of the heights
of barriers between different local minima. In Ref.~\cite{EnkO} it was
observed that there is another characteristic
local minimum around energy $-10.0$ kcal/mol
(type B in Ref.~\cite{EnkO})
whose structure is significantly different from the
groundstate structure
(type A in Ref.~\cite{EnkO}).  We took one of this type B
conformations as
the initial conformation and made a set of multicanonical
annealing runs with
100,000 Monte Carlo sweeps for various values of the sampling
interval size $\Delta E$.
  The results are presented in Table~6.  They imply
that the sampling interval size has to be more than
14 kcal/mol,
before
the system is able to find the groundstate. Hence, the barrier height
which separates the configurations of type A and type B
is between $-14$ and $-15$
kcal/mol.
Note that we can see this clearly in the results for fixed
sampling interval
size
in Table~4. For $\Delta E= 5$ and 10 kcal/mol the simulation
cannot
get  out of a local minimum once it falls in it, but
for $\Delta E = 15$
kcal/mol simulation can overcome the energy barriers and always find the
groundstate ($n_{GS} = 10$).
This kind of analysis should allow one to study in more detail
the relation between the
groundstate and long-living metastable state with
only slightly higher energy.

\subsection{Calculating Thermodynamic Quantities}
In the previous subsection, we concentrated solely on the
ability of the different algorithms
to explore the groundstate of the peptide. In this section we like to
demonstrate in more detail how the regular multicanonical method (not
multicanonical annealing) is able to
 estimate thermodynamic quantities over a wide range of temperatures.
This amazing new feature results from
the fact that it is possible to reconstruct
the canonical distribution at various temperatures from one
multicanonical simulation (see Eq.~(\ref{erw})).
 As an example we show in Fig.~1
the probability distribution as a function of the temperature that
was obtained by the multicanonical production run with 1,000,000
MC sweeps.  As expected, there is a conspicuous peak in probability
around $E=-11$ kcal/mol near $T=0$ K.
 In principle, one could also reconstruct the canonical
distribution  from
simulated annealing runs provided that the annealing schedule is
monitored very carefully.\cite{Dave2}
However, in practice this constraint is difficult to realize and
introduce hard-to-control systematic errors in the calculation
of thermodynamic quantities.
 As an example we show in
  Fig.~2a and Fig.~2b the probability distribution at $T=300$ K
  and $T = 50$ K as obtained by a canonical simulation, a multicanonical
  simulation, and
   simulated annealing. All simulations
   relied on 1,000,000 sweeps.
In the cases of canonical simulation and simulated annealing, our
   results are from 20 runs of 50,000 sweeps with different
 random start-configurations.  Only the final conformations were
taken into account for these cases, and so the results are very
crude estimate of the probability distribution.  For the multicanonical
case, on the other hand, not only the final conformation but also all
other intermediate conformations were taken
into account through reweighting.
   For 300 K the multicanonical simulation reproduced
  the distribution as obtained directly by a canonical distribution while
  simulated annealing gives a much rougher approximation.
   For 50 K the canonical simulation got trapped in a local
  minimum and did not reach the equilibrium and is far away from the
   the true distribution as predicted by the multicanonical
  simulation. But
  simulated annealing also failed in reproducing this
distribution. The center of
  the two distributions (obtained by multicanonical simulation
and simulated annealing)
  differ by $\approx 2$ kcal/mol.
  This demonstrates
  clearly that calculating thermodynamic quantities from
  simulated annealing runs
  is dangerous and requires a careful monitoring of the annealing to
    provide reliable results. On the other hand,
  multicanonical simulations are not hindered by this problem and we can
  use them
   to calculate a variety of physical quantities.

An important quantity one likes to monitor as a function of
temperature is
the average energy of the peptide. This energy is a sum of four terms:
Coulomb-energy $E_{es}$, hydrogen-bond energy $E_{hb}$,
van der Waals energy
$E_{vdW}$ and torsion energy $E_{tor}$ (see Eqns.~(16)--(20)).
In Fig.~3
we display expectation values of these terms and the total energy
$E_{tot}$
 as a function of the temperature. One clearly observes
that the behavior
of the energy $E_{tot}$ is dominated by the van der Waals term.
\cite{Yold}
While the behavior of the energy is smooth and slowly changing,
we observe a pronounced peak in
 the specific heat, defined by
\begin{equation}
  C(\bhat)  = {\bhat}^2 \frac{<E^2> - <E>^2}{5}
\end{equation}
(see Fig.~4), a phenomena common to
phase transitions in statistical physics. An important question is if the
peak is indeed a signal for a phase transition.
 Of course our peptide is a finite system and
much too
small to speak of a phase transition in the sense the word is
used in statistical physics, but one can imagine that the peak indicates a
crossover between a ``folded'' and a random coil structure which could
for larger proteins be interpreted as a phase transition.

To study this question one first has to find a  quantity which can serve as an
order parameter.
A natural definition
of such an order parameter is inspired by the Parisi
order parameter for spin glasses: the overlap of two configurations
separated by an infinitely long time:
\begin{equation}
Op= \frac{1}{N_{\alpha}}\sum_{\alpha}
\cos ({\alpha} (t) - {\alpha} (t+\infty)),
\end{equation}
where the sum goes over all $N_{\alpha}$
dihedral angles $\alpha \in ( \phi,\psi,\chi ) $ (here,
$N_{\alpha}=19$).
This definition is by no means unique.
For instance one could also choose
\begin{equation}
Op = \frac{1}{N_{\alpha}}
         \sum_{\alpha} \delta ({\alpha} (t), {\alpha} (t+\infty)),
\end{equation}
where $\delta(\alpha_i,\alpha_j) = 1$, if $|\alpha_i -\alpha_j| < D$ and
0 in all other cases, with $D$ a certain upper bound (say, 20 degrees).
Another way to calculate our ``order parameter''
 would be to perform two simulations of the same
 system with different start configurations
(replicas in the language of spin glasses):
\begin{equation}
Op = \frac{1}{N_{\alpha}} \sum_{\alpha}
\cos (\alpha^{(1)} - \alpha^{(2)}),
\end{equation}
where the superscript marks the different replicas. Since both
replicas are
totally uncorrelated they can be regarded as the same system
separated by an infinitely long time.
We remark that similar order parameters for the
system of a biomolecule were considered by other workers, too.
\cite{Shak,Dave2}

In the present work, we take the definition of Eq.~(23) for the
order parameter.
We approximated the requirement of infinitely long separation of the
configurations by taking the overlap of the first 100,000 sweeps after
thermalization with the last 100,000,
so that both sets are separated by 750,000 sweeps.  This is
much more than necessary,
since every tunneling event marks a new set of statistically independent
configurations, and we had 19 tunneling events altogether (see
Table~1).
Fig.~5 displays the average of the order parameter as
a function of the  temperature $T$.
 It seems that our ``order parameter'' allows
to describe in a general way the crossover between the folded state and the
coil state.
The step-like
behavior of this quantity should become even steeper for a system with
a phase transition. However, unlike for spinglasses
one always finds only
one state and not a multitude of different groundstates for low temperatures.
This can be seen in
Fig.~6 where we display
 the order parameter distribution as a function of
temperature. Note that even for high temperatures the mean of the distribution
is not at zero as for spin systems. This is due to the fact that
geometric constraints make certain angular  values  highly unfavorable.

In order to further elucidate the behavior of the order parameter
we now define a
configuration as groundstate-like if  14 (or 74 \%)
 of the dihedral angles differ
by less than 20 degrees from the lowest-energy configuration ever
encountered (
type A in Ref.~\cite{EnkO}).  We likewise define a coil
configuration as one where less than 5 of the dihedral angles fullfil the
above condition. All other configurations are considered
as intermediate.
Fig.~7 displays the percentage of the different types of
configurations as a function of temperature.  The percentage of groundstate
configurations is maximal for low temperatures and  decreases fast around
room temperatures. On  other hand, the high temperature region is dominated
by coil structures, while the intermediate structures occur mostly
at temperatures around 300 K. We also studied how often the
configurations
of types B and C of Ref.~\cite{EnkO} appear,
but found that these
configurations have no significant contributions. Only for temperatures
around 250 K we found, that about 2 \% of the configurations were of
type B, otherwise their number was always negligible.
To study in more detail
the mechanism of the crossover between the ordered structure
and the random structure we investigate
the free energy differences $\Delta G$, enthalpy differences
$\Delta H$ and
entropy differences $\Delta S$ between groundstate-like
configurations and
coil structures.  The enthalpy difference was estimated by the
difference in potential energy.  First, we display in Fig.~8
the enthalpy differences $\Delta H$.
We find that the enthalpy differences between intermediate
configurations and
coil configurations are much smaller than those between groundstate
configurations and intermediate configurations or between groundstate
configurations and coil structures. Over the whole energy range the
groundstate is energetically favored, but this is different for the
free energy or entropy.
The free energy differences were calculated from
\begin{equation}
\Delta G = - {\beta}^{-1} \ln { {N_{A}} \over {N_{B}}} ,
\end{equation}
where $N_{A}$ and $N_{B}$ are the average numbers of configurations in
states $A$ and $B$, respectively.
Finally, the entropy differences were obtained from
\begin{equation}
T\Delta S = \Delta H - \Delta G~.
\end{equation}
As one can see from Fig.~9 coil structures are favored by entropy.
Again the differences between
intermediate configurations and coil structures are much smaller than
between groundstates and coils or between groundstates and intermediate
structures. For temperatures below room temperature the entropy difference
between intermediate configurations and coil configurations  vanishes.
 For temperatures below 150 K the entropy difference between groundstates
and coil configuration is also compatible with zero.
Since intermediate states and coil structures have similar entropy and energy
we treat them now as the same type of configurations and  compare
the groundstate configurations with non-groundstate configurations.
In Fig.~10 we display the
enthalpy, and entropy differences between both types of
configurations.
Again one can see that the thermodynamic behavior of the
peptide is dominated
for high temperatures by the entropy,
 favoring non-groundstate-like structures. Hence, we observe
large positive
free energy differences $\Delta G$. On the other hand at
low temperatures energy dominates and
 favors groundstate-like structures, yielding negative free
energy differences $\Delta G$.  Around room temperature $\Delta G$ is
small, groundstate-like configurations and others appear with similar
probability. This explains the large fluctuations in the total energy
visible in the peak in the specific heat.

\section{CONCLUSIONS AND DISCUSSION}
By performing a simulation with Met-enkephalin, we studied the
performances of a few variants of the multicanonical
approach and compared them with simulated annealing. It
was shown that the multicanonical
method allows a more efficient search for the groundstate.
Even
more important, multicanonical simulation allows one to study the thermodynamic
behavior of biological molecules over a wide range of temperatures from just
one simulation, which was not possible by other methods. We demonstrated this
by calculating various physical quantities as a function of temperature
and estimated  differences in the free energy, enthalpy, and
entropy between
groundstate-like configurations and other configurations.

\vspace{0.5cm}
\noindent
{\Large \bf Acknowledgements}:\\
Our simulations were
performed on  clusters of fast RISC workstations at SCRI (The Florida
State University, Tallahassee, USA) and
HITAC S820/80 at National Institute for Molecular Science, Okazaki,
Japan.
This work is
supported, in part, by the Department of Energy, contract DE-FC05-85ER2500,
by a Grant-in-Aid for Scientific Research from the
Japanese Ministry for Education, Science and Culture, by the
Konrad-Zuse-Zentrum f{\"u}r Informationstechnik Berlin (ZIB) and by MK
Industries, Inc. Part of this work was written while one of us (U.~H.)
was summer visitor at Brookhaven National Laboratory. U.~H.\ likes to
thank BNL for the kind hospitality extended to him.\\


\newpage
\noindent
{\Large Table Captions:}\\
{Table~1: Estimated groundstate energies $E_{GS}$ (in kcal/mol)
of each tunneling event
as obtained by
a regular multicanonical simulation. $t_{min}$
is the sweep when the simulation first entered the groundstate region ($E \le
-11.0$ kcal/mol) in the corresponding tunneling event.
$<E>$ is the average of these
estimates $E_{GS}$.}\\
{Table~2: Lowest energy (in kcal/mol) obtained by simulated annealing
for three different choices of $NSTEP$ and two different
choices of the final temperature $T_f$, where $NSTEP$ is the number
of different temperatures considered in the annealing process
(see Eqns.~(\ref{ana})--(\ref{anc})).
For all cases, the total number of Monte Carlo steps per run, $NSWEEP$,
was 100,000.
$<E>$ is the average of the lowest energy estimates and $n_{GS}$
the number of runs in which a conformation with $E \le -11.0$ kcal/mol
was obtained.}\\
{Table~3: Lowest energy (in kcal/mol) obtained by simulated annealing
with $NSTEP=NSWEEP$, where $NSTEP$ is the number
of different temperatures considered in the annealing process
and $NSWEEP$ is the total number of Monte Carlo steps per run.
Here, we have $NSWEEP=50,000$.
Two different
choices of the final temperature $T_f$ ($=50$ K and 1 K) were
considered.
$<E>$ is the average of the lowest energy estimates and $n_{GS}$
the number of runs in which a conformation with $E \le -11.0$ kcal/mol
was obtained.} \\
{Table~4: Lowest energy (in kcal/mol) obtained by multicanonical annealing
for different choices of the annealing sampling size $\Delta E$.
$\Delta_m$
is the minimal sampling size in the case where $\Delta E$ was changed
dynamically.
For all cases, the total number of Monte Carlo steps per run, $NSWEEP$,
was 100,000.
$<E>$ is the average of the lowest energy estimates and $n_{GS}$
the number of runs in which a conformation with $E \le -11.0$ kcal/mol
was
obtained.}\\
{Table~5: Lowest energy (in kcal/mol) obtained by multicanonical
annealing
for the best choices of the annealing sampling size $\Delta E$ and
$\Delta_m$ in Table~4, where $\Delta_m$
is the minimal sampling size in the case where $\Delta E$ was changed
dynamically.
For all cases, the total number of Monte Carlo steps per run, $NSWEEP$,
was 50,000.
$<E>$ is the average of the lowest energy estimates and $n_{GS}$
the number of runs in which a conformation with $E \le -11.0$ kcal/mol
was
obtained.}\\
{Table~6: Type of lowest-energy structure and its
energy $E_m$ (in kcal/mol)
obtained by multicanonical annealing runs with various
sampling interval size
$\Delta E$. The initial conformation for each run was that of a
local minimum
state (type B structure defined in Ref.~\cite{EnkO}). Type A is a
groundstate structure and has $E \le -11.0$ kcal/mol.}\\

\newpage
Table~1.\\
\begin{table}[h]
\begin{center}
\begin{tabular}{||c|c|c||}\hline \hline
$n_{tu}$ & $t_{min}$ & $E_{GS}$\\ \hline
 0 & 22104 & $-12.0$ \\
 1 & 78808 & $-11.9$ \\
 2 & 123478 & $-12.0$ \\
 3 & 211464 & $-12.0$ \\
 4 & 267742 & $-12.1$ \\
 5 & 299228 & $-12.1$ \\
 6 & 372020 & $-11.6$ \\
 7 & 375092 & $-12.1$ \\
 8 & 445356 & $-12.1$ \\
 9 & 499032 & $-12.0$ \\
10 & 553208 & $-11.9$ \\
11 & 569618 & $-11.4$ \\
12 & 655420 & $-12.0$ \\
13 & 673768 & $-12.0$ \\
14 & 781366 & $-11.3$ \\
15 & 796264 & $-11.9$ \\
16 & 875756 & $-12.0$ \\
17 & 971704 & $-11.8$ \\
18 & 977694 & $-12.1$ \\ \hline
$<E> $ & ---   & $-11.9(2)$ \\ \hline \hline
\end{tabular}
\end{center}
\label{tab0}
\end{table}

\newpage
Table~2.\\
\begin{table}[h]
\begin{center}
{\small
\begin{tabular}{||c||c|c||c|c||c|c||}\hline \hline
& \multicolumn{2}{c}{$NSTEP=100,000$}&
  \multicolumn{2}{c}{$NSTEP=50$}&
  \multicolumn{2}{c}{$NSTEP=20$}\\ \hline
Run&$T_f = 50$ K&$T_f = 1$ K&$T_f = 50$ K&$T_f = 1$ K&
    $T_f = 50$ K&$T_f = 1$ K
\\ \hline
1  & $-11.9$ & $-8.0$ & $-12.0$ & $-11.2$ & $-11.9$ & $-10.9$ \\
2  & $-11.6$ & $-11.9$ & $-11.9$ & $-9.1$ & $-7.5$ & $-11.8$ \\
3  & $-9.2$ & $-8.5$ & $-10.0$ & $-8.3$ & $-10.8$ & $-9.0$ \\
4  & $-7.8$ & $-8.0$ & $-8.2$ & $-10.0$ & $-10.6$ & $-12.2$ \\
5  & $-7.3$ & $-12.2$ & $-9.5$ & $-12.2$ & $-8.2$ & $-7.9$ \\
6  & $-12.0$ & $-9.8$ & $-12.0$ & $-11.2$ & $-10.9$ & $-10.0$ \\
7  & $-10.9$ & $-10.3$ & $-11.8$ & $-10.6$ & $-11.9$ & $-9.1$ \\
8  & $-12.0$ & $-11.1$ & $-11.0$ & $-12.2$ & $-9.4$ & $-7.1$ \\
9  & $-12.0$ & $-10.3$ & $-9.3$ & $-8.6$ & $-11.9$ & $-8.5$ \\
10 & $-9.0$ & $-8.6$ & $-8.2$ & $-12.2$ & $-10.3$ & $-9.0$ \\ \hline
$<E>$ & $-10.4(1.9)$ & $-9.9(1.6)$ & $-10.4(1.5)$ & $-10.6(1.5)$ &
$-10.3(1.5)$ & $-9.6(1.7)$ \\ \hline
$n_{GS}$& 5/10 & 3/10 & 5/10 & 5/10 & 3/10 & 2/10 \\ \hline \hline
\end{tabular}
}
\end{center}
\label{tab1}
\end{table}

\newpage
Table~3.\\
\begin{table}[h]
\begin{center}
{\small
\begin{tabular}{||c||c|c||}\hline \hline
Run&$T_f = 50$ K&$T_f = 1$ K
\\ \hline
1  & $-10.5$ & $-11.7$ \\
2  & $-10.6$ & $-8.6$ \\
3  & $-11.6$ & $-12.1$ \\
4  & $-8.7$ & $-8.8$ \\
5  & $-9.3$ & $-7.4$ \\
6  & $-8.7$ & $-8.9$ \\
7  & $-8.3$ & $-12.1$ \\
8  & $-8.3$ & $-12.2$ \\
9  & $-8.8$ & $-7.1$ \\
10 & $-8.2$ & $-7.5$ \\
11 & $-11.9$ & $-9.9$ \\
12 & $-10.3$ & $-7.3$ \\
13 & $-11.8$ & $-8.4$ \\
14 & $-9.5$ & $-10.6$ \\
15 & $-11.7$ & $-10.3$ \\
16 & $-8.6$ & $-12.2$ \\
17 & $-10.5$ & $-12.2$ \\
18 & $-10.2$ & $-9.1$ \\
19 & $-11.6$ & $-11.9$ \\
20 & $-11.8$ & $-12.1$ \\ \hline
$<E>$ & $-10.0(1.4)$ & $-10.0(2.1)$ \\ \hline
$n_{GS}$& 6/20 & 8/20 \\ \hline \hline
\end{tabular}
}
\end{center}
\label{tab1b}
\end{table}

\newpage
Table~4.\\
\begin{table}[h]
\begin{center}
{\small
\begin{tabular}{||c||c|c|c||c|c||}\hline \hline
& \multicolumn{3}{c}{Fixed sampling size}&
  \multicolumn{2}{c}{Dynamically changed sampling size}\\ \hline
Run&$\Delta E = 5$ & $\Delta E = 10$ & $\Delta E =15 $
& $ \Delta_{m} = 5$ & $\Delta_{m} = 10$
\\ \hline
1  & $-8.9$ & $-11.9$ & $-12.0$ & $-10.4$ & $-11.8$ \\
2  & $-10.1$ & $-11.9$ & $-11.6$ & $-11.7$ & $-11.8$ \\
3  & $-11.8$ & $-10.5$ & $-12.0$ & $-11.6$ & $-11.8$ \\
4  & $-10.6$ & $-10.3$ & $-11.8$ & $-11.3$ & $-11.3$ \\
5  & $-9.3$ & $-10.3$ & $-12.0$ & $-11.9$ & $-11.6$ \\
6  & $-7.6$ & $-12.0$ & $-11.9$ & $-10.5$ & $-11.8$ \\
7  & $-12.0$ & $-11.8$ & $-12.0$ & $-11.6$ & $-11.5$ \\
8  & $-12.0$ & $-11.5$ & $-11.5$ & $-11.0$ & $-11.6$ \\
9  & $-9.4$ & $-10.8$ & $-11.5$ & $-11.9$ & $-11.6$ \\
10 & $-9.6$ & $-10.3$ & $-11.8$ & $-11.7$ & $-11.8$ \\ \hline
$<E>$ & $-10.1(1.5)$ & $-11.1(8)$ & $-11.8(2)$ & $-10.9(1.1)$ & $-11.7(2)$
\\ \hline
$n_{GS}$& 3/10 & 5/10 & 10/10 & 8/10 & 10/10
\\ \hline \hline
\end{tabular}
}
\end{center}
\label{tab2}
\end{table}

\newpage
Table~5.\\
\begin{table}[h]
\begin{center}
{\small
\begin{tabular}{||c||c||c||}\hline \hline
& Fixed sampling size &
  Dynamically changed sampling size \\ \hline
Run& $\Delta E =15 $ & $\Delta_{m} = 10$ \\ \hline
1  & $-11.6$ & $-11.7$ \\
2  & $-12.0$ & $-11.3$ \\
3  & $-10.2$ & $-11.5$ \\
4  & $-10.1$ & $-10.3$ \\
5  & $-11.9$ & $-11.5$ \\
6  & $-12.0$ & $-11.5$ \\
7  & $-11.9$ & $-11.1$ \\
8  & $-11.7$ & $-11.1$ \\
9  & $-11.8$ & $-11.3$ \\
10 & $-11.9$ & $-11.5$ \\
11 & $-12.0$ & $-10.1$ \\
12 & $-12.1$ & $-11.6$ \\
13 & $-12.0$ & $-11.3$ \\
14 & $-11.8$ & $-11.6$ \\
15 & $-11.3$ & $-10.4$ \\
16 & $-11.9$ & $-11.8$ \\
17 & $-12.0$ & $-10.1$ \\
18 & $-11.9$ & $-11.8$ \\
19 & $-11.9$ & $-10.8$ \\
20 & $-11.6$ & $-11.5$ \\ \hline
$<E>$ & $-11.7(6)$ & $-11.2(6)$ \\ \hline
$n_{GS}$& 18/20 & 15/20
\\ \hline \hline
\end{tabular}
}
\end{center}
\label{tab2b}
\end{table}

\newpage
Table~6.\\
\begin{table}[h]
\begin{center}
\begin{tabular}{||c|c|c||}\hline \hline
$\Delta E$ & $E_m$ & Type of structure \\ \hline
10 & $-10.5$ & B\\
11 & $-10.8$ & B\\
12 & $-10.5$ & B\\
13 & $-10.5$ & B \\
14 & $-10.6$ & B\\
15 & $-11.5$ & A \\ \hline \hline
\end{tabular}
\end{center}
\label{tab3}
\end{table}
\newpage
\noindent
{\Large Figure Captions:}\\
{Fig.~1: Probability distribution $P (E)$ of the energy
as a function of temperature $T$ obtained
by a regular multicanonical simulation with 1,000,000 MC sweeps.}\\
 {Fig.~2a: Probability distribution of the energy for T = 300 K as obtained by
 a canonical simulation ({\it Cano}), a multicanonical simulation ({\it MuCa})
 and simulated annealing ({\it SiAn}).}\\
 {Fig.~2b: Probability distribution of the energy for T = 50 K as obtained by
 a canonical simulation ({\it Cano}), a multicanonical simulation ({\it MuCa})
 and simulated annealing ({\it SiAn}).}\\
{Fig.~3: Average total energy
$E_{tot} = E_{es}+E_{HB}+E_{vdW}+E_{tor}$ ($+$)
and averages  of its component terms, Coulomb energy $E_{es}$ ($X$),
hydrogen-bond energy $E_{HB}$ ($\diamondsuit$),
 van der Waals energy $E_{vdW}$ ($\Box$), and torsion energy
$E_{tor}$ ($\bigcirc$)
 as a function of temperature $T$. All values are calculated
from a regular
 multicanonical simulation with 1,000,000 MC sweeps.}\\
{Fig.~4: Specific heat as a function of
temperature $T$.
The values are calculated from a regular
 multicanonical simulation with 1,000,000 MC sweeps.}\\
{Fig.~5: Average of the order parameter $Op$ defined in Eq.~(23)
  as a function of temperature $T$.  The values are calculated
from a regular
 multicanonical simulation with 1,000,000 MC sweeps.}\\
{Fig.~6: Probability distribution of the order parameter $Op$
(defined in
 Eq.~(23)) as a function of temperature $T$.
The values are calculated from a regular
 multicanonical simulation with 1,000,000 MC sweeps.}\\
{Fig.~7: Fraction of groundstate-like configurations ({\it GS}), coil
configurations ({\it COIL}), and
intermediate configurations ({\it IM}) as a function of
temperature $T$. A groundstate-like configuration differs no more than
$20 $ degrees in at least 14 of the 19 dihedral angles from the
groundstate;
a coil configuration
differs in at least 14 dihedral angles by more than 20 degrees from the
groundstate configuration.  The values are calculated from a regular
 multicanonical simulation with 1,000,000 MC sweeps.}\\
{Fig.~8: Enthalpy differences $\Delta H$
   between groundstate-like configurations and coil configurations ($+$),
   between groundstate-like configurations and intermediate configurations
   ($\bigcirc$), and between intermediate configurations and coil
   configurations ($X$) as a function of temperature $T$.
The values are calculated from a regular
 multicanonical simulation with 1,000,000 MC sweeps.}\\
{Fig.~9: Entropy differences $T \Delta S$
   between groundstate-like configurations and coil configurations ($+$),
   between groundstate-like configurations and intermediate configurations
   ($\bigcirc$), and between intermediate configurations and coil
   configurations ($X$) as a function of temperature $T$.
The values are calculated from a regular
 multicanonical simulation with 1,000,000 MC sweeps.  The errors were
   smaller than the plot symbols.}\\
{Fig.~10: Free energy differences $\Delta G$ ($\triangle$), enthalpy
     differences $\Delta H$ ($+$)
 and entropy differences $T \Delta S$ ($\Box$)
 between groundstate-like configurations
 and non-groundstate-like configurations
 as a function of temperature $T$.
The values are calculated from a regular
 multicanonical simulation with 1,000,000 MC sweeps.
The errors were smaller than
  the plot symbols.}\\
\end{document}